\documentclass[lOpt]{article} 
\usepackage{amsmath}
\usepackage{amssymb}
\usepackage{float} 
\usepackage[T1]{fontenc} 
\usepackage{graphicx} 
\usepackage[utf8]{inputenc} 
\usepackage{mathtools} 
\usepackage{txfonts} 
\usepackage{wasysym} 
\usepackage[backend=biber, sorting=none]{biblatex}  \usepackage[paperheight=29.7cm,paperwidth=21.0cm,left=2.5cm,right=2.5cm,top=2.5cm,bottom=2.0cm]{geometry} 

\title{Patterning Silver Nanowire Network via the Gibbs—Thomson Effect} 

\begin{document} 
\maketitle 

\author{
    Hongteng Wang$^{a}$, 
    Haichuan Li$^{a}$, 
    Yijia Xin$^{a}$, 
    Weizhen Chen$^{b}$, 
    Haogen Liu$^{a}$, 
    Ying Chen$^{a}$, 
    Yaofei Chen$^{a,c}$, 
    Lei Chen$^{a,c*}$, 
    Yunhan Luo$^{a,c*}$, 
    Zhe Chen$^{a,c}$, 
    and Gui-Shi Liu$^{a,c*}$
}

$\ast$ Corresponding author. 

Email: chenlei@jnu.edu.cn; yunhanluo@163.com; guishiliu@163.com

\vspace{1\baselineskip}
\textbf{Abstract}: As transparent electrodes, patterned silver nanowire (AgNW) networks suffer from noticeable pattern visibility, which is an unsettled issue for practical applications such as display. Here, we introduce a Gibbs-Thomson effect (GTE)-based patterning method to effectively reduce pattern visibility. Unlike conventional top-down and bottom-up strategies that rely on selective etching, removal, or deposition of AgNWs, our approach focuses on fragmenting nanowires primarily at the junctions through the GTE. This is realized by modifying AgNWs with a compound of diphenyliodonium nitrate and silver nitrate, which aggregates into nanoparticles at the junctions of AgNWs. These nanoparticles can boost the fragmentation of nanowires at the junctions under an ultralow temperature (75°C), allow pattern transfer through a photolithographic masking operation, and enhance plasmonic welding during UV exposure. The resultant patterned electrodes have trivial differences in transmittance ($\Delta $T = 1.4\%) and haze ($\Delta $H = 0.3\%) between conductive and insulative regions, with high-resolution patterning size down to 10 $\mu$m. To demonstrate the practicality of this novel method, we constructed a highly transparent, optoelectrical interactive tactile e-skin using the patterned AgNW electrodes.


\vspace{1\baselineskip}

\newpage

\section{Introduction}

Flexible transparent electrodes (TEs) are essential components in various optoelectronic devices\textsuperscript {1}, including solar cells\textsuperscript {2-4}, touch sensors\textsuperscript {5, 6}, light-emitting diodes\textsuperscript{7-9}, transparent strain/pressure sensors\textsuperscript{10-13}, and wearable sensing\textsuperscript {14, 15}. Silver nanowire (AgNW) is a prominent material for flexible TEs due to their exceptional optoelectronic properties, excellent mechanical flexibility, and easy processing\textsuperscript {16-18}. AgNW networks need to be assembled in micro-patterns to construct functional devices or integrated electronics\textsuperscript {7,17,19}. For the applications in transparent electronics, special attention should be paid to the optical visibility of AgNW patterns due to the nontrivial light scattering and absorption of AgNW near the UV band.

There are two patterning strategies for AgNWs: (1) top-down methods that initially deposit an AgNW network and then selectively remove a part of the network, such as photolithography\textsuperscript{20, 21} and etching\textsuperscript{22}; (2) bottom-up methods that directly assemble AgNWs into micropatterns, such as self-assembly\textsuperscript{23, 24} and ink printing\textsuperscript{25-29}; Regardless of the patterning methods used, the patterned TEs generally do not contain AgNWs in insulative regions17, yielding the differences in scattering and adsorption between the insulative and conductive regions. The optical difference leads to optical traces in the AgNW patterns\textsuperscript{30}. The traces are undesirable for transparent electronics due to issues like degradation of display quality. Reshaping pentagonal NW into circular NW can reduce the adsorption of surface plasmonic resonance to lower the optical traces\textsuperscript{31}, which is not suitable for flexible TE due to the high light power involved. Reducing the electrode linewidth to <10 $\mu$m is another approach to creating visually imperceptible AgNW patterns7. However, such narrow AgNW electrodes have drawbacks of highly unstable conductivities among inter-electrodes or even disconnection\textsuperscript{32}. Maintaining the nanowire morphology within the insulative region can create an optical match in terms of scattering and adsorption throughout the AgNW pattern, which inherently eliminates the pattern traces. There is a patterning method based on Plateau-Rayleigh instability that meets the above requirements\textsuperscript{33}. However, this method requires a high-temperature annealing (193°C), higher than glass transition temperatures of common flexible substrates, which limits its applications in flexible electronics.

Here, we report a Gibbs-Thomson effect (GTE)-based patterning method for AgNW to overcome the issue of optical invisibility. The GTE method is different from both the conventional “top-down” and “bottom-up” strategies which generally involve etching/removing and selective deposition to realize patterning, respectively. It achieves AgNW patterns through the GTE-induced fragmentation of AgNWs at the junctions at a low temperature, which retains nanowire fragments to eliminate the optical mismatch between the conductive and insulative regions. The fragmentation is facilitated by the composition of diphenyliodonium nitrate (DPIN) and silver nitrate (DA), which is well dissolved in the AgNW dispersion. The DA can be self-assembled into nanoparticles (NPs) mainly at the junctions of AgNWs, which function as a photoresist for pattern transfer, a solder for plasmonic welding, and a fluxing agent for fragmentation of AgNWs simultaneously. The fabricated TE has welded AgNW networks, high-resolution patterns, and excellent optical invisibility. The GTE-based method only requires UV exposure and thermal annealing at low temperatures (down to 75°C) to form high-resolution AgNW patterns, which is straightforward and scalable for high throughput manufacturing.
\section{Experimental Section}

\subsection{Materials}

AgNO\textsubscript{3} solution (0.1 mol/L) and DPIN powder were purchased from Sigma-Aldrich Inc. Ethanol-dispersed AgNWs (10 mg/mL) were obtained from Zhejiang Ke Chuang Inc. DPIN, acetone, and deionized water were mixed at a weight ratio of 1:12.5:12.5. The AgNO\textsubscript{3} solution was added into the DPIN mixture at a weight ratio of 3:1 to form the DA solution. The DA solution was further blended with the diluted AgNW solution (2 mg L-1) to form the DA-AgNW ink for use. The PDMS was prepared as follows: PDMS prepolymer and curer (SYLGARD™ 184 silicone kit) were mixed at a weight ratio of 10:1. The mixture was thoroughly stirred and defoamed in a vacuum tank for 40 minutes for use. 
\subsection{Patterning of DA-AgNWs}
The defoamed PDMS prepolymer was spin-coated on a cleaned glass or flexible substrate, followed by curing at 100°C for 3 hours. Then, an amount of DA-AgNW solution was spin-coated on the PDMS/glass that had been cleaned by air plasma (Sunjune Plasma VP-R5). Afterward, the dried DA-AgNW network was covered by a photomask and exposed to UV light (center wavelength: 365 nm, CEAULIGHT CEL-HXF300) for pattern transfer. Lastly, the AgNW substrate was annealed on a hot plate for 3 minutes at temperatures ranging from 75 to 135°C.
\subsection{Preparation of the optoelectronic interactive tactile system} 
The AgNW electrode array with the diamond-shaped pattern was fabricated on a PDMS film with a thickness of 100 $\mu$m. The silver wires were affixed to the ends of the electrode array using silver paste. The two electrode films peeled from supporting substrates, were laminated together using PDMS as a glue. After thermal curing, a soft projected capacitive sensor is constructed. The silver wire leads are connected to two ESP32 microcontrollers. Utilizing the ESP-NOW protocol, the ESP32 units can communicate with each other. Finally, a connection between the ESP32 and the LED array is established via the gpio14 pin to complete the tactile system.
\subsection{Optical Simulation}
FDTD simulations were performed using Ansys Lumerical Solution software. 2D FDTD was conducted for the simulation of electric field intensity distribution and 3D FDTD was conducted for the simulation of photothermal generation. The AgNW with a pentagonal cross-section was built by AutoCAD software for the simulations (Fig. S3). The dielectric function of Ag is from the literature\textsuperscript{34}. The plane wave was used for the simulation of electric field intensity distribution and the total field scattered field source was employed for absorption and scattering simulation. To simulate infinite NW length, PML was chosen as the boundary condition. A non-uniform meshing strategy was implemented, featuring a refined grid with a resolution of 0.3 nm near the AgNW junction to precisely capture plasmonic interactions. In contrast, a coarser mesh was applied to other regions to enhance computational efficiency while maintaining accuracy.
\subsection{Characterization}
The morphologies of the AgNWs and electrode patterns were characterized using an SEM (Carl Zeiss SUPRA 60) and/or an optical microscope (Carl Zeiss Axio Scope.A1). The XRD spectra of the AgNWs were determined using an X-ray polycrystalline diffractometer (Bruker AXS GmbH D8 Advance). The transmittance and haze spectra were measured using a UV-visible spectrophotometer (Shimadzu UV-3600 plus). Surface modifications of AgNWs were characterized using an X-ray Photoelectron Spectroscopy (XPS) (Thermo Scientific K-Alpha X). 

\vspace{1\baselineskip}
\section{Results and Discussion}

The invisible AgNW pattern is achieved through selective fragmentation of an AgNW network via the Gibbs-Thomson effect at a low temperature (down to 75°C). Fig. 1a-c illustrate the GTE-based patterning procedure for AgNWs. This procedure, while similar to conventional photolithography, eliminates the requirement for photoresist and physical/chemical etching. We ingeniously developed an electronic ink comprising AgNWs and DA, where the DA serves triple roles: it acts as a photoresist-like layer for patterning transfer, as a fluxing agent for micromachining the AgNW network, and as a solder for plasmonic welding AgNWs. The procedure begins with spin-coating the AgNW ink on a substrate. Upon drying, DA self-assembles into NPs and selectively deposits at the junctions of the AgNWs (Fig. 1e and 1h). Subsequently, the DA-decorated AgNW film (DA-AgNWs) undergoes UV exposure through a photomask to trigger selectively decomposition of the DA while simultaneously weld the junctions between AgNWs via the plasmonic resonance effect. The AgNW network with the DA pattern is then annealed at a low temperature to form a AgNW patterned AgNW through the GTE-induced fragmentation, mainly at the junctions (Fig. 1g, 1j). The GTE-based patterning largely preserves the morphology of the nanowire network in the insulative region, ensuring that the optical properties (scattering and adsorption) remain consistent throughout the patterned electrode (Fig. 1k). As a result, the GTE-patterned AgNW electrodes exhibit excellent optical invisibility, which is unparalleled by other conventional patterning methods such as photolithography, as depicted in Fig. 1d, 1k.

\subsection{GTE-based fragmentation of DA-AgNW networks}
DA nanoparticles tend to aggregate at junctions of the AgNWs (Fig. 1h). These DA nanoparticles facilitate the fragmentation of AgNWs at the junctions during thermal annealing and promote welding under UV exposure (Fig. 2a). The selective DA deposition can be ascribed to the capillary effect at the junctions\textsuperscript{35}. During spin coating of the AgNW ink, the centrifugal force removes excess liquid, leaving a small amount of ink at junctions due to the capillary effect (Fig. 2b). In addition, because of the difference in surface curvatures, the junctions have low chemical potentials36, which also directs the DA deposition towards the intersections. The deposition process is affected by coating methods and ink properties (Fig. S1). An optimized DA self-assembly is obtained using spin-coating with a DA solution at pH 4.6 and a mass fraction of 0.63\%. 

\begin {figure} [H]
\centerline{\includegraphics{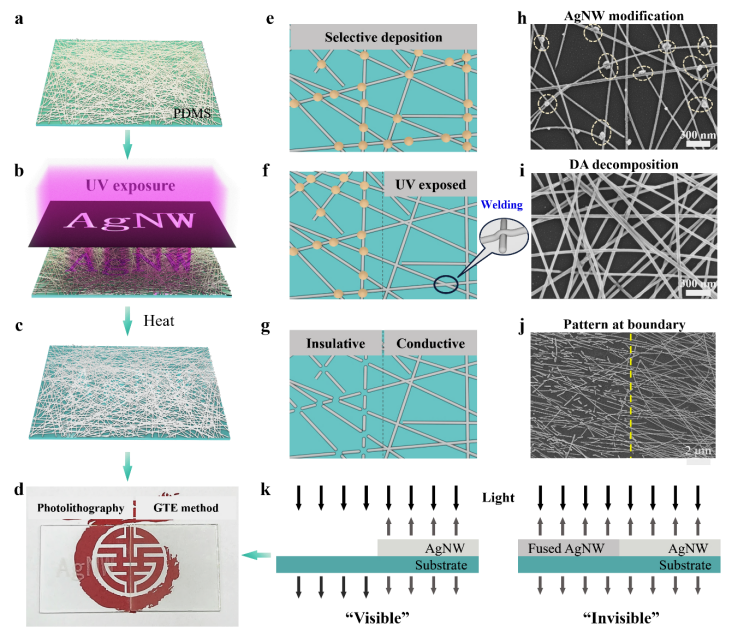}}
\caption{GTE-based patterning of AgNWs. (a-c) Schematics of the patterning procedure. (d) Photographs of the photolithography-processed (left) and GTE-induced (right) AgNW patterns. Schematics of the (e) selective deposition of DA at the junctions of AgNWs, (f) half-UV-exposed DA-AgNWs, and (g) half-fragmented AgNW network. Scanning electron microscope (SEM) image of (h) the DA-AgNWs, (i) UV-exposed DA-AgNWs, and (j) patterned DA-AgNWs. (k) Comparison of the conventional and the GTE-induced AgNW patterns in the aspect of optical visibility.}
\label {fig:1}
\end {figure}

The DA modification can form Ag-based molten salts on the AgNW surface, which significantly reduces the fusing temperature of the AgNW networks. As illustrated in Fig. 2c, X-ray diffraction (XRD) analysis of the DA-AgNWs reveals the formation of three distinct silver compounds on the AgNW surface: AgNO\textsubscript{3}, AgIO\textsubscript{3}, and AgI. The characteristic peaks for AgNO\textsubscript{3} are observed at 33.43°, 49.07°, 61.98°; 29.60° and 47.91° corresponding to AgIO\textsubscript{3}; and 43.78° assigned to AgI. The XPS spectra further confirms the presence of NO\textsubscript{3}\textsuperscript{-}, IO\textsubscript{3}\textsuperscript{-}, and I- on the DA-AgNWs surface. The N 1s spectra reveal a peak associated with NO\textsubscript{3}\textsuperscript{-}, indicating the oxidation of AgNW in the presence of DA (Fig. S2). Similarly, the I 3d spectra exhibit signals corresponding to IO\textsubscript{3}\textsuperscript{-} and I-, further validating the formation of AgIO\textsubscript{3} and AgI (Fig. 2d). After UV irradiation, the I 3d signal significantly diminishes, suggesting DA decomposition and subsequent chemical transformations. Following low-temperature annealing, a shift in the I 3d peaks indicates redox interactions, further contributing to the fusion of AgNWs. These compounds have lower melting points (211°C for AgNO\textsubscript{3}, ~200°C for AgIO\textsubscript{3}, 552°C for AgI) than that of raw AgNW (~236°C for 30 nm diameter)\textsuperscript{36}. The three compounds at a specific ratio can form a molten salt with an ultralow melting point (76°C)\textsuperscript{37}. Therefore, there is an optimal proportion of diphenyliodonium nitrate (DPIN) to AgNO\textsubscript{3} to form the molten salt. The optimal weight ratio of DPIN to AgNO\textsubscript{3} is found to be 1:3, which produces the lowest melting point of the DA-AgNW network (Fig. S3).

The DA nanoparticles contribute to the fragmentation of the AgNW networks at the junctions under a low temperature. It is reported that the Ag-based molten salt expedites the surface atom diffusion and triggers the PRI of AgNW\textsuperscript{38}. Here, since the DA is mainly deposited at the junctions, the fragmentation is dominated by the Gibbs-Thomson effect, instead of the Plateau-Rayleigh instability effect. The curvature differential between the surface of the nanowires and their intersections creates a gradient in chemical potential. This gradient induces a net diffusion (\(J_{net}\)) toward the junction at elevated temperature (i.e., the Gibbs-Thomson effect)\textsuperscript{38}:

\begin {figure} [H]
\centerline{\includegraphics{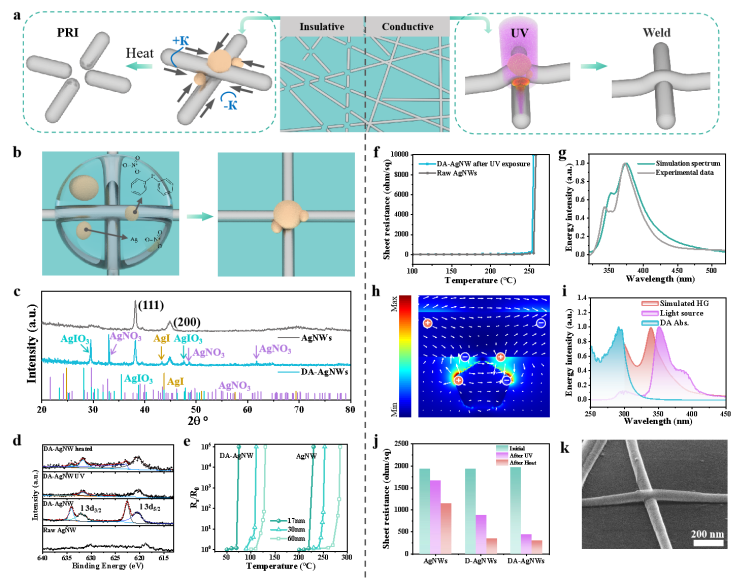}}
\caption{(a) Schematic of the DA-based fragmentation and welding of an AgNW network. (b) Schematic of selective deposition of DA at the junction of two stacked AgNWs. (c) XRD spectra of raw AgNWs and DA-AgNWs. (d) High-resolution XPS spectra of the I 3d region for raw AgNW, DA-AgNWs, UV-treated DA-AgNWs, and heated DA-AgNWs. (e) \textit{R\textsubscript{s}/R\textsubscript{0} } of raw AgNWs and DA-AgNWs with different average diameters during annealing at different temperatures. (f) \textit{R\textsubscript{s} } change of the UV-exposed DA-AgNWs and raw AgNWs after annealing at different temperatures for 3 minutes. (g) Simulated extinction spectra of a pentagonal AgNW and experimental extinction spectrum of AgNWs. (h) Simulated electric field distribution near two stacked AgNWs at 365 nm wavelength, the white arrows indicate the vectorial electric field. (i) Absorption spectrum of DA, spectrum of the UV source, and the HG spectrum obtained by an FDTD simulation. (j) \textit{R\textsubscript{s} } change of raw AgNW, D-AgNW, and DA-AgNWs networks before and after UV and UV \& heating treatment. (k) SEM image of the welded AgNWs. }
\label {fig:2}
\end {figure}

\vspace{1\baselineskip}
\begin{equation}
J_{\text{net}} = - \left( \frac{D_s \gamma \Omega v}{kT} \right) (\kappa_j - \kappa_{\text{AgNW}}) e
\end{equation}

where \(K\) is the Boltzmann constant, \(D_{s}\) is the surface diffusion coefficient, \(\gamma\) denotes surface tension, \(\Omega\) is the atomic volume, \(\nu\) is the number of diffusing atoms per unit surface area, \textit{T} is the temperature, \(\kappa_{J}\) is the local mean curvature of the junction, \(\kappa_{AgNW}\) is the mean curvature of AgNW, and \textbf{\textit{e}} is the axial unit vector. The DA nanoparticles at the junctions act as both fluxing agents and perturbation points of curvatures, which can further promote this diffusion. Therefore, the DA-AgNW network can trigger the fragmentation of the AgNW network at the junctions with low temperatures (Fig. S4). To quantify the fluxing effect, the temperature at \textit{R\textsubscript{s}/R\textsubscript{0} }> 10\textsuperscript {5} is defined as fusing temperature \textit{T\textsubscript{f}}. As shown in Fig. 2e, the DA reduces the \textit{T\textsubscript{f}} by an average of 150.6$^{\circ}C$ for the AgNWs with mean diameters in the range of 17 nm to 90 nm. The \textit{T\textsubscript{f}} already decreases to 75$^{\circ}C$, which is lower than the glass transition temperatures of common flexible substrates for fabricating flexible electronics.

\begin {figure} [H]
\centerline{\includegraphics{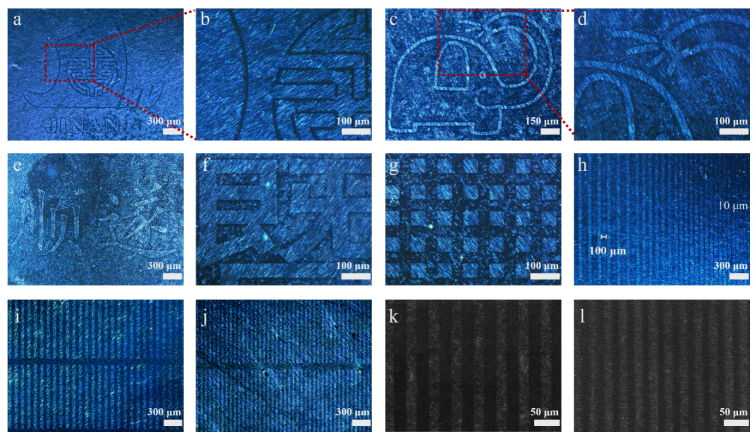}}
\caption{High-resolution patterns of DA-AgNWs. (a-j) OM images of the patterned DA-AgNW networks with different graphic designs. The side length of the squares in (g) is 50 $\mu$m. (h) The pattern with a fixed linewidth of 50 $\mu$m and spacings gradually reduced from 100 to 10 $\mu$m. The linewidths of (i-j) are 50 $\mu$m and 30 $\mu$m, respectively. (k)-(l) SEM images of the AgNW patterns with linewidths of 20 $\mu$m and 10 $\mu$m, respectively. }
\label {fig:3}
\end {figure}

\subsection{Plasmonic welding of DA-AgNWs}
The fluxing effect of DA also promotes the plasmonic welding of AgNWs under light exposure. Since the characteristic extinction peaks of AgNWs are located at around 350 nm and 374 nm (Fig. 2g), UV light is used to weld the AgNWs. The two peaks can be ascribed to the excitation of the transverse plasmon resonance. Finite-difference time-domain (FDTD) simulations indicate that the cylindrical nanowire only yields one peak with small deviation from the experimental spectrum (Fig. S5). By contrast, two nondegenerate quadrupolar modes occur for the pentagonal twinned nanowire\textsuperscript{39}, which outputs the simulated extinction spectrum well matched with the experimental result (Fig. 2g). Therefore, the pentagon-shaped AgNW is selected as the simulation model thereafter. The dielectric function of silver was obtained from the literature34 and fitted in FDTD (Fig. S6), resulting in simulated spectra that exhibit well agreement with the experimentally measured spectra of AgNW (Fig. S5). At the junction of two stacked AgNWs, the UV shining can produce strong electromagnetic coupling between the longitudinal plasmon and the local surface plasmon in the two nanowires (Fig. S7, Fig. 2h). This coupling amplifies the field strength by several orders of magnitude and induces the heat pot due to the collective oscillation of electrons\textsuperscript{40}. The wavelength has a great impact on the coupling and thus the localized photothermal effect (Fig. S7-S8). We quantify the heat generation (HG) near the junction using the Joule heating equation\textsuperscript{41}:

\vspace{1\baselineskip}
\begin{equation}
q = \frac{\omega}{2} \operatorname{Im}(\varepsilon) |E|^2
\end{equation}
                
where \(q\) represents the volumetric heat source density within the nanowires, \(\omega\)  is the angular frequency of the external field, \(\varepsilon\)  is the dielectric constant in air, and \textit{E}E is the electric field within the AgNWs. The simulated spectrum indicates that UV light induces higher HG and the maximal efficiency is located at around 340 nm. We chose the UV band with a peak of 350 nm as the light source, which has a large overlap with the HG spectrum to improve plasmonic welding and a slight overlap with the absorption spectrum of DA to decompose it. The impact of UV irradiation is influenced by both exposure duration and light intensity (Fig. S9). Irradiation for 8 minutes at a light power density of 10.74 mW cm\textsuperscript{-2} can completely decompose DA. Fig. 2j and 2k shows the boost of DA to the plasmonic welding. After the same low-level UV exposure (10.74 mW cm\textsuperscript{-2}, 8 min), the sheet resistance (\textit{R\textsubscript{s} }) of raw AgNWs only decreases by 13.5\%, while the AgNW network modified solely with DPIN (D-AgNW) has a larger decrease in \textit{R\textsubscript{s} } (by 54\%) and the DA-AgNWs exhibit the largest decrease of \textit{R\textsubscript{s} }, i.e. by 77.5\%. Further low-temperature annealing induces a reduction in the \textit{R\textsubscript{s} } of all three samples, with DA-AgNWs exhibiting the most significant decrease of 84.4\%.
\subsection{Patterning of AgNWs and optical invisibility}
The UV exposure induces plasmonic welding and decomposition of DA in the meantime, which enables pattern transfer. UV light can split DPIN into phenol, iodobenzene, and nitric acid, and decompose AgNO\textsubscript{3} into Ag and nitric acid. Except for Ag, the other compounds will evaporate. SEM images indicate that trace amounts of DA particles remain on the NW surface after 6-minute UV exposure (Fig. S10) and no DA NPs are observed after 8-minute treatment (Fig. 1i). The UV-exposed DA-AgNWs recovers the thermal stability and its Tf is similar to that of raw AgNWs. The Tf difference between the exposed and shadowed AgNWs can facilely produce high-resolution patterning after a low-temperature annealing (down to 75°C). Optical microscopy (OM) images in Fig. 3 illustrate that the AgNW patterns have sharp, well-defined edges and exceptional fine-line detailing (Fig. 3b-d). Fig. 3h displays the pattern with a fixed linewidth of 50 $\mu$m and a spacing gradient ranging from 100 $\mu$m to 10 $\mu$m. The AgNW patterns with linewidths and spacing of 50, 30, 20, and 10 $\mu$m are also successfully achieved (Fig. 3i-l), demonstrating the high-resolution patterning ability of the GTE method. The GTE high-resolution mode has been successfully applied to PDMS, glass, silicon wafer, and PET substrates (Fig. S11), exhibiting high reproducibility (Fig. S12).

\begin {figure} [H]
\centerline{\includegraphics{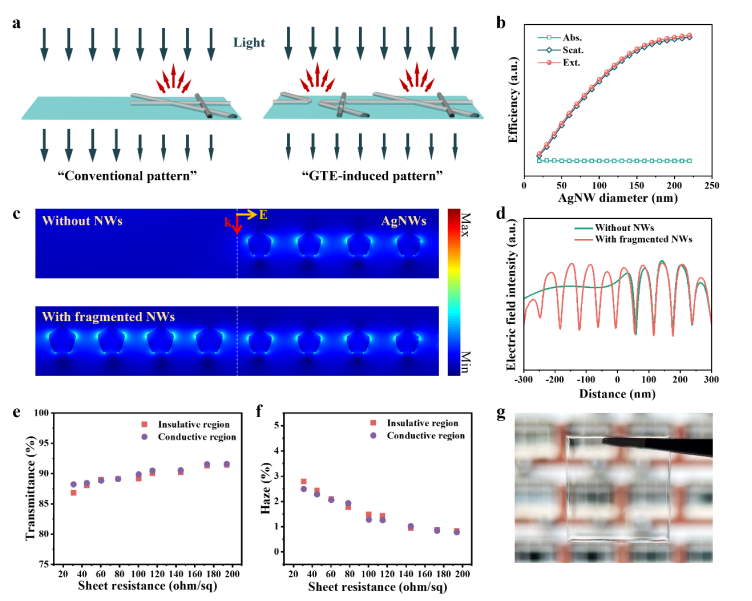}}
\caption{Pattern Invisibility. (a) Differences in light passing through a substrate for a conventional AgNW pattern and the GTE-induced AgNW pattern. (b) Simulated absorption, scattering, and extinction efficiency of AgNWs with different diameters. (c) Simulated electric field intensity distribution of the conventional and GTE-induced AgNW patterns. (d) Profiles of the electric field intensity at 10 nm below the AgNWs extracted from (c). (e) Transmittance and (f) haze difference between the conductive and insulative regions of the DA-AgNW patterns with different R\textsubscript{S}. (g) Photograph of the patterned DA-AgNW electrode.}
\label {fig:4}
\end {figure}

The DA-AgNW patterns fabricated by the GTE method are optically invisible to the naked eyes. The AgNW patterns achieved by conventional techniques generally have no nanowires in the insulative regions (Fig. 4a). This region therefore has high transmittance and low haze. In the conductive region, the AgNWs have considerable scattering ability against visible light (Fig. S13) to produce lower transmittance and higher haze\textsuperscript{42}. Such optical difference makes the conventional AgNW patterns visible to the naked eyes (i.e., pattern traces). In contrast, the GTE method can retain NW fragments in the insulative region to eliminate the optical difference throughout the AgNW pattern. The differences in optical transparency ($\Delta $T), haze ($\Delta $H), and reflectance (Fig. S14) are measured to be 1.4\%, 0.3\%, and 0.3\% between the insulative and conductive regions with \textit{R\textsubscript{s} } in the range of 31 to 145 ohm/sq, respectively (Fig. 4e, f). The $\Delta $T and $\Delta $H would increase to 11.8\% and 2.5\% in the conventional patterns respectively, since the insulative regions have no nanowires and thus 100\% transmittance and 0 haze. FDTD simulations of the electric field distribution near a boundary of patterned AgNWs display the discrepancy between the conventional and our AgNW patterns (Fig. 4c-d). The transmitted electric field distribution of the AgNW network is higher in the insulative region without nanowires, while, for our AgNW pattern, the two regions exhibit similar profiles of electric field intensity. The scattering of the fragmented nanowires is slightly stronger, since the average diameter of the fragmented AgNWs is marginally thicker than that of the AgNWs in the conductive region (Fig. S13, S15). This is the reason that the H of the insulative regions is slightly higher than those of the conductive regions (Fig. 4f). The DA-AgNWs exhibit excellent bending stability, retaining their performance after 1000 bending cycles (Fig. S16). To evaluate the oxidation resistance of DA-AgNW, the samples were stored at 85°C and 85\% relative humidity (RH) for 12 days. Only minor degradation in their electrical and optical properties was observed.

The GTE patterning technique offers significant advantages, including process simplicity, high patterning accuracy, and, most notably, the optical invisibility effect. It consists of only three fundamental steps: coating, exposure, and annealing. Compared to conventional photolithography, GTE can be completed with a simpler process, at lower annealing temperatures, and in shorter durations. While inkjet printing involves fewer steps, its resolution is limited by the nozzle size (Table S1) and it cannot overcome the issue of the optical visibility.
\subsection{Applications}
An optoelectronic interactive haptic system was constructed using the patterned DA-AgNW electrodes. The system consists of a tactile sensor, two ESP32 microcontroller chips, and an LED panel (Fig. 5a). The tactile sensor is composed of two orthogonal DA-AgNW patterns and a thin PDMS dielectric layer in between. Each electrode layer is interfaced with an ESP32 microcontroller chip. 5-row electrodes (X) and 5-column electrodes (Y) with a rhombus pattern intersect with the dielectric layer, forming a 5 × 5 cross-grid projected capacitive tactile sensor array with 25 pixels, which can attach well to the arm’s skin. When a conductive object such as a finger approaches the target point, it disturbs the electromagnetic field coupling by forming a new capacitance (Cp) and thus increases the electrode's existing capacitance to the ground. The chip scans the capacitance of each x and y electrode relative to the ground, identifying the intersection of the x and y axes as the touchpoint. The ESP32 microcontroller sends the data to an LED panel, which lights up the corresponding LED to identify the touch point. As shown in Fig. 5d, when point (2, 2) was touched, only the capacitances of the electrode X2 and Y2 changed significantly, thus the LED at (2, 2) could be lightened up. Furthermore, the device sensitively responds to various touch interactions during sliding gestures (Supporting video S1). Fig. 5e details the response of each capacitor at different touch positions and Fig. 5f shows the capacitive heatmap when a finger touches the coordinate (2,2), demonstrating the system's sensitivity and precision. The electrode demonstrates excellent mechanical stability and durability. It retains conductivity under 10\% stretching (Fig. S18a). With TPU encapsulation, stable electrical performance is maintained even after 1000 cycles of 5\% stretching, whereas without encapsulation, R/R0 gradually increases under the same conditions (Fig. S18b).

A high-sensitivity strain sensor based on stretchable serpentine electrodes was developed using the GTE method. The sensor exhibits a strain range exceeding 15\% and a high gauge factor (GF) (224 at 0.5\% strain, 400 at 15\% strain) (Fig. S19a, b). It can detects subtle physiological strains, such as joint flexion (Fig. S19c, Supporting Video S2). A 1000-cycle stretching-releasing test at 5\% strain confirms stable electromechanical performance (Fig. S19d), demonstrating its excellent mechanical durability for applications in human motion monitoring.

\begin {figure} [H]
\centerline{\includegraphics{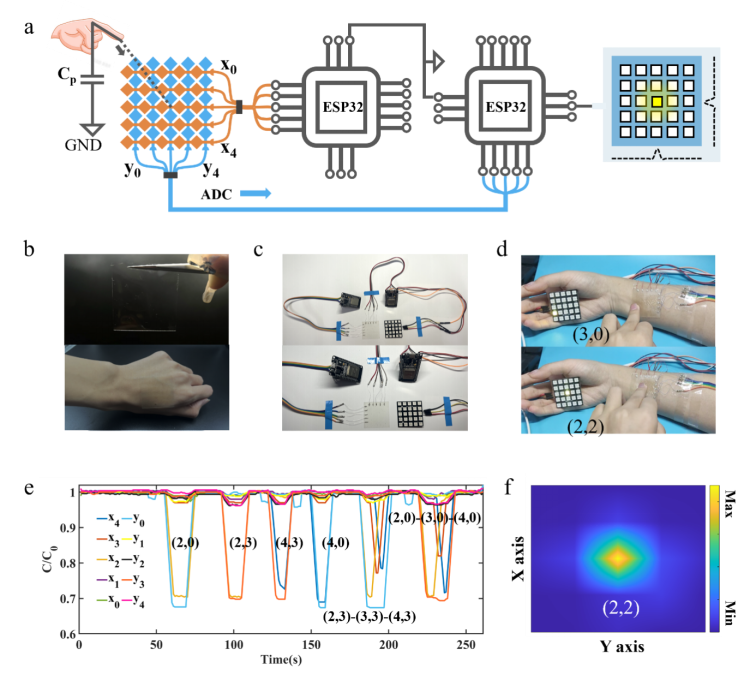}}
\caption{Tactile system based on the DA-AgNW patterns. (a) Schematic of the optoelectronic interactive tactile system. Photographs of (b) the tactile sensor on the skin, (c) the system, and (d) working demonstration of touch-responsive LED lighting. (e) Relative capacitance variation of the sensor array in response to single-pixel touch and sliding gestures along a row or column of pixels. (f) Capacitance map of $\Delta $C\textsubscript{XY} /C\textsubscript{X0Y0} induced by touching the position (2,2) showing the local capacitance variation during a single-point touch. }
\label {fig:5}
\end {figure}

\section{Conclusion}
In summary, we develop a novel Gibbs-Thomson effect (GTE)-based patterning method to fabricate AgNW patterns without optical traces. The GTE method is realized using an AgNW ink containing the DA. The DA remarkably decreases the fusing temperature of AgNW networks by 150°C and can be decomposed by UV light. Therefore, the GTE-based patterning only involves two steps: photolithographic UV exposure and thermal annealing. Different from the previous AgNW patterns, the GTE-induced AgNW pattern retains the nanowire traces in the insulative region, which remarkably eliminates the optical difference throughout the patterned electrode. The GTE method demonstrates a new route for AgNW patterning beyond the conventional top-down and bottom-up strategies. Besides pattern invisibility, this method has the merits of simple processing, high resolution, low processing temperature, and good scalability. By optimizing the light source size and collimation characteristics, the GTE method enables the easy fabrication of high-resolution patterned electrodes and large-area manufacturing processes.

\textbf{Acknowledgements}

We thank R. F. Hamans and A. Baldi for their assistance in FDTD simulations. This work was supported by the Basic and Applied Basic Research Foundation of Guangdong province (2024A1515030155, 2022A1515010272, 2024A1515012609, 2023A1515011459), National Natural Science Foundation of China (61904067, 62475101, 62175094, 62275109), open funding from the State Key Laboratory of Optoelectronic Materials and Technologies (Sun Yat-Sen University, OEMT-2022-KF-08), and Fundamental Research Funds for the Central Universities (11621405).

\textbf{Author details}

{\large\textsuperscript{a}}College of Physical \& Optoelectronic Engineering, Jinan University, Guangzhou 510632, China. {\large\textsuperscript{b}}Guangdong Provincial Key Laboratory of Optical Fiber Sensing and Communications, Key Laboratory of Visible Light Communications of Guangzhou, Key Laboratory of Optoelectronic Information and Sensing Technologies of Guangdong Higher Education Institutes, College of Physical \& Optoelectronic Engineering, Jinan University, Guangzhou 510632, China.

\textbf{Author contributions}

Hongteng Wang: Formal analysis, Data curation, Visualization, Writing – original draft. Haichuan Li: Data curation, Visualization. Yijia Xin: Data curation, Visualization. Weizhen Chen: Data curation. Haogen Liu: Data curation. Ying Chen: Data curation, Visualization. Yaofei Chen: Resources. Chen, Lei: Resources, Writing – review \& editing. Yunhan Luo: Resources. Zhe Chen: Resources. Gui-Shi Liu: Conceptualization, Formal analysis, Data curation, Resources, Writing – review \& editing.

\textbf{Competing interests}

The authors declare no competing interests.

\textbf{References}

1.	Meng, L. et al. Solution-processed flexible transparent electrodes for printable electronics. ACS Nano 17, 4180-4192 (2023).

2.	Xie, C., Liu, Y., Wei, W. \& Zhou, Y. Large‐Area Flexible Organic Solar Cells with a Robust Silver Nanowire‐Polymer Composite as Transparent Top Electrode. Adv. Funct. Mater. 33, 2210675 (2023).

3.	Yu, C. et al. Silicon solar cell with undoped tin oxide transparent electrode. Nat. Energy 8, 1119-1125 (2023).

4.	Li, M. et al. Synergistic Enhancement of Efficient Perovskite/Quantum Dot Tandem Solar Cells Based on Transparent Electrode and Band Alignment Engineering. Adv. Energy Mater., 2400219 (2024).

5.	Han, C. et al. Flexible Tactile Sensors for 3D Force Detection. Nano Lett. 24, 5277-5283 (2024).

6.	Chen, W. et al. Ionic Liquid Enabled Transparent Heterogeneous Elastomers for Soft Electronics. Adv. Funct. Mater., 2405002 (2024).

7.	Zhang, K. et al. An Ultra‐High Transparent Electrode via a Unique Micro‐Patterned AgNWs Crossing‐Network with 3.9\% Coverage: Toward Highly‐Transparent Flexible QLEDs. Adv. Funct. Mater. 34, 2308468 (2024).

8.	Du, M. et al. Facile Nanowelding Process for Silver Nanowire Electrodes Toward High‐Performance Large‐Area Flexible Organic Light‐Emitting Diodes. Adv. Funct. Mater., 2404567 (2024).

9.	Yao, L. Q. et al. High‐efficiency stretchable organic light‐emitting diodes based on ultra‐flexible printed embedded metal composite electrodes. InfoMat 5, e12410 (2023).

10.	Liu, G.-S. et al. Ultrasonically patterning silver nanowire–acrylate composite for highly sensitive and transparent strain sensors based on parallel cracks. ACS Appl. Mater. Interfaces 12, 47729-47738 (2020).

11.	Chen, L. et al. Flexible and transparent electronic skin sensor with sensing capabilities for pressure, temperature, and humidity. ACS Appl. Mater. Interfaces 15, 24923-24932 (2023).

12.	Lei, Y. et al. Surface engineering AgNW transparent conductive films for triboelectric nanogenerator and self-powered pressure sensor. Chem. Eng. J. 462, 142170 (2023).

13.	Wang, T. et al. High Sensitivity, Wide Linear‐Range Strain Sensor Based on MXene/AgNW Composite Film with Hierarchical Microcrack. Small 19, 2304033 (2023).

14.	Kong, L. et al. Wireless Technologies in Flexible and Wearable Sensing: from Materials Design, System Integration to Applications. Adv. Mater., 2400333 (2024).

15.	Liu, H. et al. Machine-Learning Mental-Fatigue-Measuring $\mu$m-Thick Elastic Epidermal Electronics (MMMEEE). Nano Lett. 24, 16221-16230 (2024).

16.	Son, D. et al. An integrated self-healable electronic skin system fabricated via dynamic reconstruction of a nanostructured conducting network. Nat. Nanotechnol. 13, 1057-1065 (2018).

17.	Chen, Y. et al. Self-assembly, alignment, and patterning of metal nanowires. Nanoscale Horiz. 7, 1299-1339, doi:10.1039/d2nh00313a (2022).

18.	Hu, S. et al. High-performance fiber plasmonic sensor by engineering the dispersion of hyperbolic metamaterials composed of Ag/TiO 2. Opt. Express 28, 25562-25573 (2020).

19.	Yang, J. et al. Multilayer ordered silver nanowire network films by self‐driven climbing for large‐area flexible optoelectronic devices. InfoMat, e12529 (2024).

20.	Zhang, W. et al. Structural multi-colour invisible inks with submicron 4D printing of shape memory polymers. Nat. Commun. 12, 112 (2021).

21.	Tian, X. et al. Crosslinking-induced patterning of MOFs by direct photo-and electron-beam lithography. Nat. Commun. 15, 2920 (2024).

22.	Cao, K. et al. Dual-mode SPR/SERS biosensor utilizing metal nanogratings fabricated via wet etching-assisted direct laser interference patterning. Appl. Surf. Sci., 159621 (2024).

23.	Zhang, X.-R. et al. Patterned nanoparticle arrays fabricated using liquid film rupture self-assembly. Langmuir 39, 10660-10669 (2023).

24.	Hu, Y., Yang, D. \& Huang, S. Simple and ultrafast fabrication of invisible photonic prints with reconfigurable patterns. Adv. Opt. Mater. 8, 1901541 (2020).

25.	Liu, H. et al. Large‐Area Flexible Perovskite Light‐Emitting Diodes Enabled by Inkjet Printing. Adv. Mater. 36, 2309921 (2024).

26.	Prakoso, S. P., Li, Y. T., Lai, J. Y. \& Chiu, Y. C. Concept of Photoactive Invisible Inks toward Ultralow‐Cost Fabrication of Transistor Photomemories. Adv. Electron. Mater. 9, 2201147 (2023).

27.	Mazzotta, A. et al. Invisible thermoplasmonic indium tin oxide nanoparticle ink for anti-counterfeiting applications. ACS Appl. Mater. Interfaces 14, 35276-35286 (2022).

28.	Li, W. et al. Microsecond-scale transient thermal sensing enabled by flexible Mo1- xWxS2 alloys. Research 7, 0452 (2024).

29.	Li, W. et al. Large‐scale ultra‐robust MoS2 patterns directly synthesized on polymer substrate for flexible sensing electronics. Adv. Mater. 35, 2207447 (2023).

30.	Tomiyama, T., Mukai, I., Yamazaki, H. \& Takeda, Y. Optical properties of silver nanowire/polymer composite films: absorption, scattering, and color difference. Optical Materials Express 10, 3202-3214 (2020).

31.	Hwang, J., Lee, H. \& Woo, Y. Enhancing the optical properties of silver nanowire transparent conducting electrodes by the modification of nanowire cross-section using ultra-violet illumination. Journal of Applied Physics 120 (2016).

32.	Yang, B.-R. et al. Microchannel wetting for controllable patterning and alignment of silver nanowire with high resolution. ACS Appl. Mater. Interfaces 7, 21433-21441 (2015).

33.	Liu, G.-S. et al. Self-assembled monolayer modulated Plateau-Rayleigh instability and enhanced chemical stability of silver nanowire for invisibly patterned, stable transparent electrodes. Nano Res. 15, 4552-4562 (2022).

34.	Yang, H. U. et al. Optical dielectric function of silver. Phys. Rev. B 91, 235137 (2015).

35.	Liu, Y. et al. Capillary-force-induced cold welding in silver-nanowire-based flexible transparent electrodes. Nano Lett. 17, 1090-1096 (2017).

36.	Liu, G.-S. et al. Comprehensive stability improvement of silver nanowire networks via self-assembled mercapto inhibitors. ACS Appl. Mater. Interfaces 10, 37699-37708 (2018).

37.	Patil, K., Rao, C., Lacksonen, J. \& Dryden, C. The silver nitrate-iodine reaction: Iodine nitrate as the reaction intermediate. J. Inorg. Nucl. Chem. 29, 407-412 (1967).

38.	Liu, G.-S. et al. One-step plasmonic welding and photolithographic patterning of silver nanowire network by UV-programable surface atom diffusion. Nano Res., 1-10 (2022).

39.	Hamans, R. F., Parente, M., Garcia-Etxarri, A. \& Baldi, A. Optical properties of colloidal silver nanowires. J. Phys. Chem. C 126, 8703-8709 (2022).

40.	Bastús, N. G., Piella, J. \& Puntes, V. Quantifying the sensitivity of multipolar (dipolar, quadrupolar, and octapolar) surface plasmon resonances in silver nanoparticles: The effect of size, composition, and surface coating. Langmuir 32, 290-300 (2016).

41.	Landau, L. D. et al. Electrodynamics of continuous media. Vol. 8 (Elsevier, 2013).

42.	Fang, Y. S. et al. High-Performance Hazy Silver Nanowire Transparent Electrodes through Diameter Tailoring for Semitransparent Photovoltaics. Adv. Funct. Mater. 28, 8, doi:10.1002/adfm.201705409 (2018).
\end{document}

pdflatex mypaper.tex

\maketitle

\noindent\textbf{* Corresponding author.} 

\noindent Email: \href{mailto:chenlei@jnu.edu.cn}{chenlei@jnu.edu.cn}, 
\href{mailto:yunhanluo@163.com}{yunhanluo@163.com}, 
\href{mailto:guishiliu@163.com}{guishiliu@163.com}.

\newpage

\begin{figure}[H] % H 表示强制当前位置插入
    \centering
    \includegraphics[width=\textwidth]{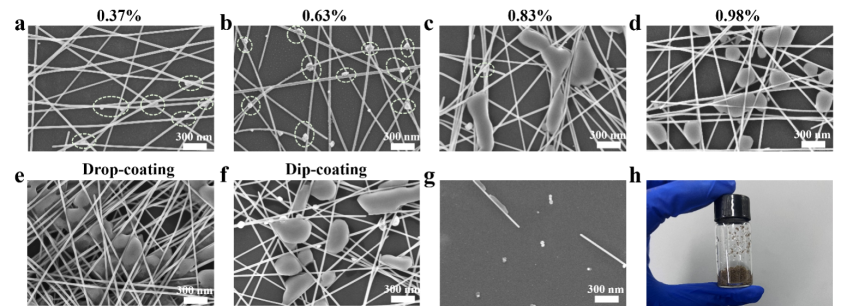} % 替换为你的图片文件
    \caption{\textbf{Fig. S1.} Self-assembly behavior of DA under different conditions. (a) - (d) SEM images of DA-AgNW solutions with mass fractions of 0.37\%, 0.63\%, 0.83\%, and 0.98\% spin-coated on PDMS substrates. SEM images of a 0.63\% DA solution deposited on the substrate via (e) drop-casting and (f) dip-coating. (g) SEM image of a 0.63\% DA solution on a PDMS substrate after nitric acid addition. (h) Photograph of precipitate formation in a 0.63\% DA solution upon NaOH addition. Spin-coating DA solutions with a 0.63\% mass fraction results in uniform surface modification of AgNWs. Lower concentrations lead to incomplete modification, while higher concentrations induce DA particle aggregation. Both drop-coating and dip-coating techniques using the optimized 0.63\% DA-AgNW solution result in DA aggregation on the substrate surfaces. The solution has a natural pH of 4.6. In acidic conditions, AgNWs undergo fragmentation and dissolution, whereas in alkaline conditions, precipitate formation occurs.}
    \label{fig:S1}
\end{figure}

\newpage

\begin{figure}[H] % H 表示强制当前位置插入
    \centering
    \includegraphics[width=\textwidth]{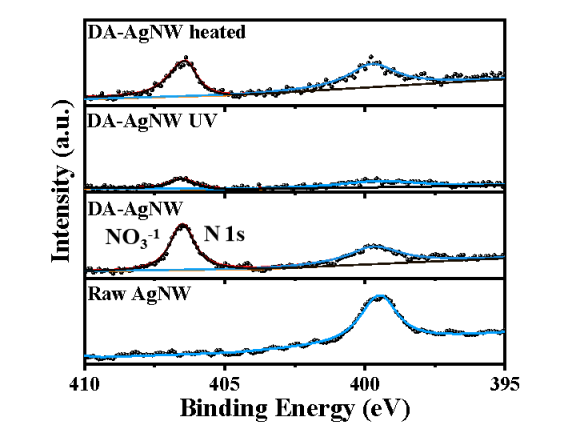} % 替换为你的图片文件
    \caption{\textbf{Fig. S2.} High-resolution XPS spectra of the N 1s region for raw AgNW, DA-AgNWs, UV-treated DA-AgNWs, and heated DA-AgNWs. In the high-resolution N 1s spectrum, a weak 400.1 eV peak, attributed to residual PVP, was detected on AgNW. After DA modification, a NO\textsubscript{3}\textsuperscript{-} peak appeared, confirming AgNO\textsubscript{3} formation. Upon UV irradiation (10.74 mW cm\textsuperscript{-2}, 8 min), the N 1s peak disappeared, indicating nitrate photolysis into volatile nitric acid. Heat treatment weakened but did not eliminate the N 1s signal, suggesting DA’s role in AgNW fusing.}
    \label{fig:S2}
\end{figure}

\newpage

\begin{figure}[H] % H 表示强制当前位置插入
    \centering
    \includegraphics[width=\textwidth]{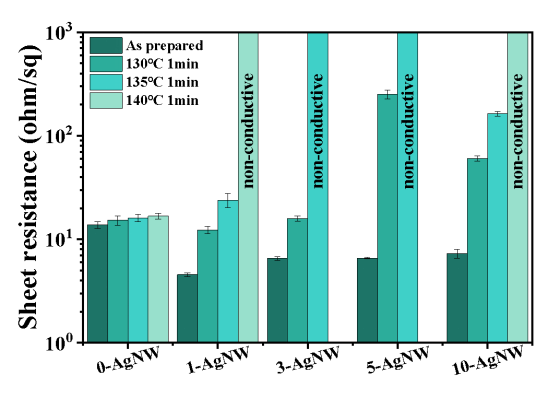} % 替换为你的图片文件
    \caption{\textbf{Fig. S3.} Variation in sheet resistance of the DA-AgNWs fabricated with different weight ratios of DA versus annealing temperature. We prepared the DA-AgNW inks with five ratios of DPIN to AgNO\textsubscript{3} (1:0, 1:1, 1:3, 1:5, 1:10) and named 0-AgNW, 1-AgNW, 3-AgNW, 5-AgNW, 10-AgNW, respectively. The DA inks with the ratio of 1:3 and 1:5 are more effective in lowering the fusing temperature (\textit{T\textsubscript{f}}) of AgNWs. The \textit{T\textsubscript{f}} of the AgNW network with a mean diameter of 60 nm is reduced from 285°C to 130°C. To avoid excess AgNO\textsubscript{3}, the 3-AgNW ink was used in our experiments.}
    \label{fig:S3}
\end{figure}

\begin{figure}[H] % H 表示强制当前位置插入
    \centering
    \includegraphics[width=\textwidth]{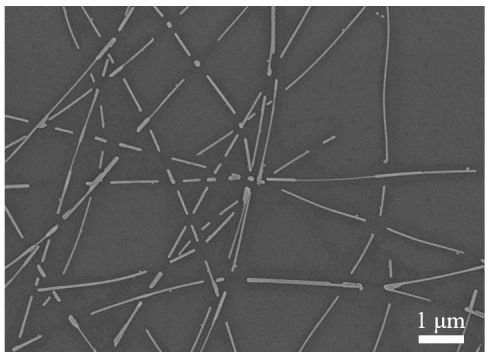} % 替换为你的图片文件
    \caption{\textbf{Fig. S4.} SEM image of the DA-AgNW network fragmented at the junctions due to the Gibbs-Thomson effect.}
    \label{fig:S4}
\end{figure}

\newpage
\begin{figure}[H] % H 表示强制当前位置插入
    \centering
    \includegraphics[width=\textwidth]{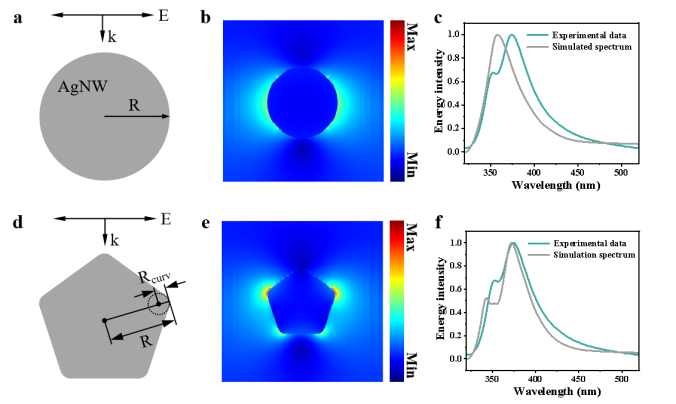} % 替换为你的图片文件
    \caption{\textbf{Fig. S5.} Optical simulation of infinitely long AgNWs with circular and pentagonal cross-sections. (a) Schematic of an infinite circular nanowire under the incident light with polarization direction perpendicular to the axial direction of NW. (b) Simulated electric field distributions for a circular nanowire with \( R = 45 \) nm. (c) Comparison between the simulated extinction spectrum of an infinite circular nanowire and the experimentally measured spectrum. (d) Schematic of an infinite pentagonal nanowire with radius \( R = 45 \) nm and radius of curvature \( R_{\text{curv}} = 5 \) nm. (e) Simulated electric field distributions for the pentagonal nanowire. (f) Comparison between the simulated extinction spectrum of the pentagonal nanowire and the experimental spectrum. (c) and (f) indicate that the AgNW with a pentagonal cross-section output better matching between FDTD simulations and experimental results.
}
    \label{fig:S5}
\end{figure}

\begin{figure}[H] % H 表示强制当前位置插入
    \centering
    \includegraphics[width=\textwidth]{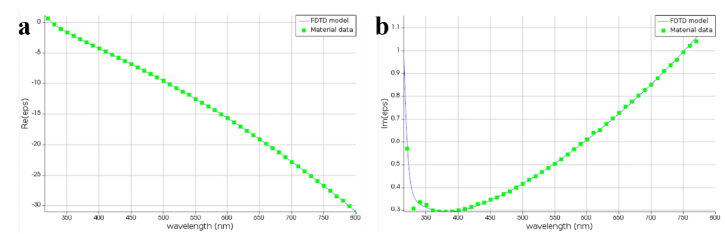} % 替换为你的图片文件
    \caption{\textbf{Fig. S6.} (a) Fitting of the real part of the dielectric function. (b) Fitting of the imaginary part of the dielectric function. The green dots denote the experimentally obtained dielectric function, while the blue curves represent the results from FDTD simulations. For the FDTD simulations, the experimental dielectric function from the literature1 is fitted.
}
    \label{fig:S6}
\end{figure}

\begin{figure}[H] % H 表示强制当前位置插入
    \centering
    \includegraphics[width=\textwidth]{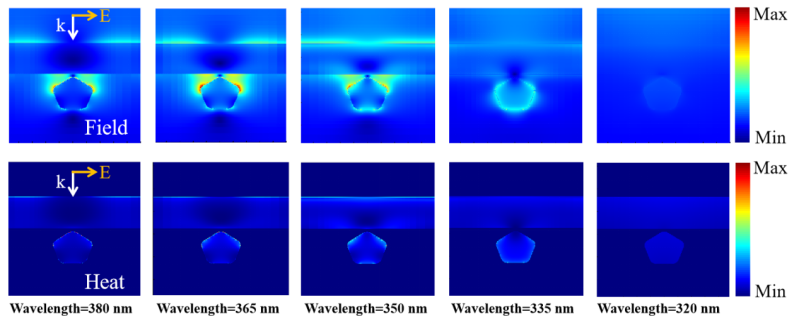} % 替换为你的图片文件
    \caption{\textbf{Fig. S7.} The local field and thermal response to light with wavelengths between 320 nm and 380 nm polarized along the top of the AgNW.
}
    \label{fig:S7}
\end{figure}

\begin{figure}[H] % H 表示强制当前位置插入
    \centering
    \includegraphics[width=\textwidth]{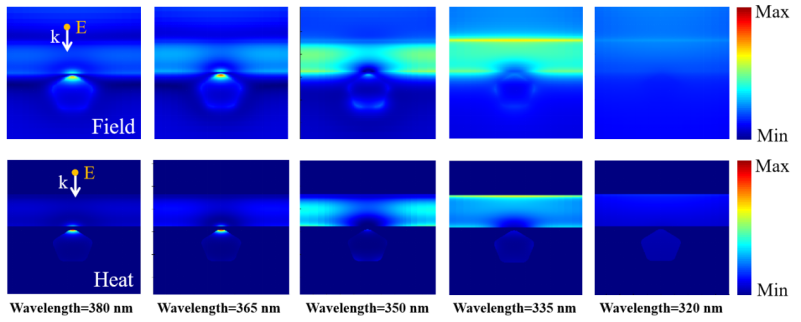} % 替换为你的图片文件
    \caption{\textbf{Fig. S8.} The local field and thermal response to light, with a wavelength ranging from 320 nm to 380 nm, polarized perpendicular to the top of the AgNW. The heat was calculated using Equation 2 in the main text. The electric field-induced heating map indicates that the most intense heat generation occurs in the gap between the two nanowires, which facilitates the welding of the AgNWs. 
}
    \label{fig:S8}
\end{figure}

\newpage
\begin{figure}[H] % H 表示强制当前位置插入
    \centering
    \includegraphics[width=\textwidth]{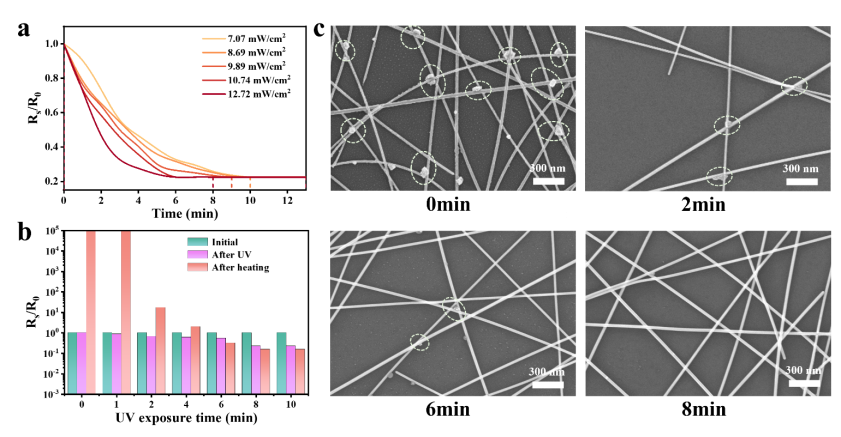} % 替换为你的图片文件
    \caption{\textbf{Fig. S9.} Effect of UV power intensity and irradiation duration on the properties of DA-AgNWs. (a) \( R_s/R_0 \) changes of DA-AgNWs under UV irradiation at different power intensities. (b) \( R_s/R_0 \) changes of DA-AgNWs under UV irradiation (\( 10.74 \, \text{mW} \, \text{cm}^{-2} \)) for different durations and after subsequent thermal treatment. (c) SEM images of DA-AgNWs under UV irradiation (\( 10.74 \, \text{mW} \, \text{cm}^{-2} \)) for different times.
}
    \label{fig:S9}
\end{figure}

\begin{figure}[H] % H 表示强制当前位置插入
    \centering
    \includegraphics[width=\textwidth]{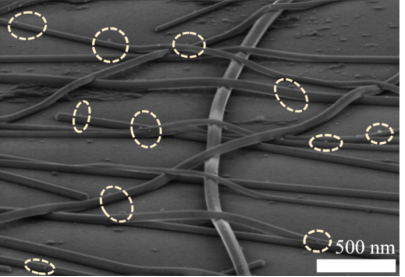} % 替换为你的图片文件
    \caption{\textbf{Fig. S10.} SEM image of the DA-AgNWs after UV exposure for 6 minutes. UV irradiation for 6 minutes is unable to achieve the complete decomposition of DA NPs, as highlighted in the yellow circles.
}
    \label{fig:S10}
\end{figure}

\newpage
\begin{figure}[H] % H 表示强制当前位置插入
    \centering
    \includegraphics[width=\textwidth]{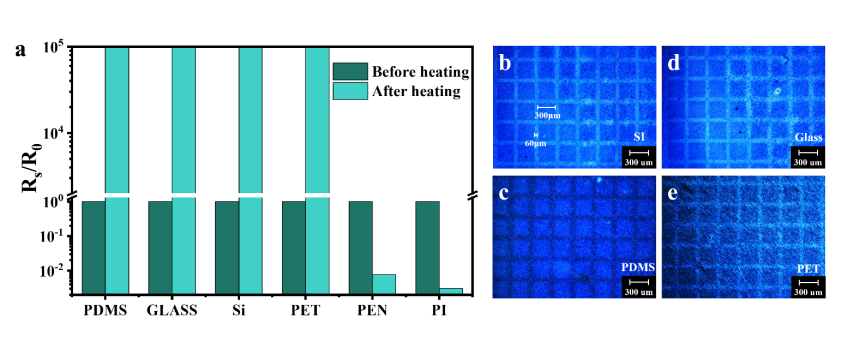} % 替换为你的图片文件
    \caption{ \textbf{Fig. S11.} Applicability of the GTE process on different substrates. (a) \( R_s/R_0 \) changes of DA-AgNWs before and after 115\(^\circ\)C heating on different substrates. (b) - (e) OM images of the DA-AgNW patterns on Si, PDMS, Glass, and PET substrates, respectively. The DA-AgNW films were deposited on six polymeric substrates: PDMS, glass, silicon, PET, PEN, and PI. A comparative analysis of \( R_s/R_0 \) before and after annealing revealed distinct substrate-dependent behaviors (Fig. S11a). After treatment, PDMS, glass, silicon, and PET substrates became completely electrically insulating (\( R_s/R_0 > 10^5 \)). However, PEN and PI substrates were incompatible with this method. Additionally, Fig. S11b-e demonstrates that DA successfully enabled precise patterning (300 \( \mu\)m edge length with 60 \( \mu\)m spacing) on Si, PDMS, glass, and PET substrates.}
    \label{fig:S11}
\end{figure}

\begin{figure}[H] % H 表示强制当前位置插入
    \centering
    \includegraphics[width=\textwidth]{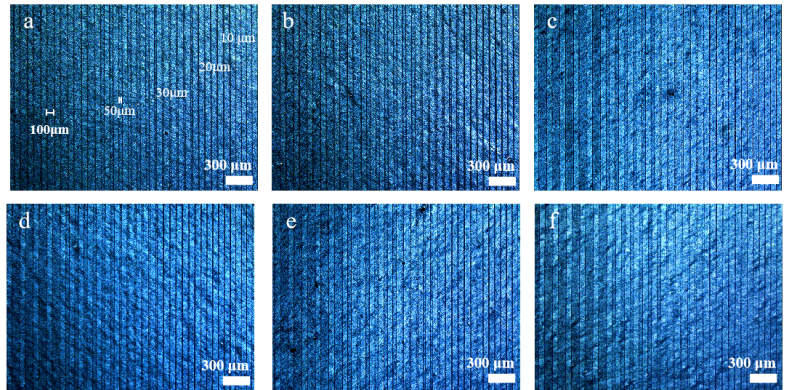} % 替换为你的图片文件
    \caption{\textbf{Fig. S12.} Reproducibility of the GTE-based patterning method. (a) - (f) OM images of the fabricated AgNW patterns with linewidth/spacing gradients ranging from 100 $\mu$m to 10 $\mu$m.
}
    \label{fig:S12}
\end{figure}

\newpage
\begin{figure}[H] % H 表示强制当前位置插入
    \centering
    \includegraphics[width=\textwidth]{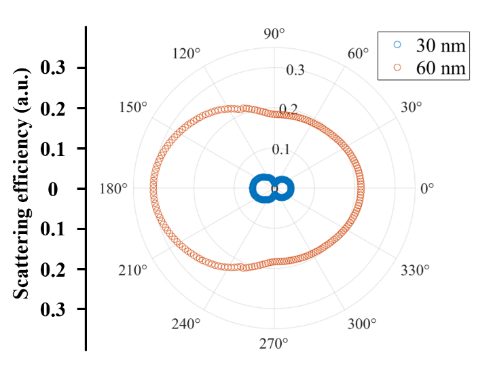} % 替换为你的图片文件
    \caption{\textbf{Fig. S13.} Simulated scattering efficiency as a function of scattering angle from the axis parallel to the incident light of 550 nm. Two diameters of 30 and 60 nm are compared. Simulations of the near-field scattering diagram indicate that forward scattering efficiency (90°-270°) predominates over backward scattering efficiency (-90° to 90°), and the scattering efficiency significantly increases with NW diameters.
}
    \label{fig:S13}
\end{figure}

\begin{figure}[H] % H 表示强制当前位置插入
    \centering
    \includegraphics[width=\textwidth]{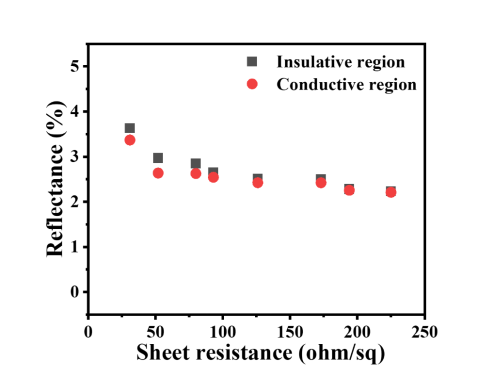} % 替换为你的图片文件
    \caption{\textbf{Fig. S14.} Reflectivity at a wavelength of 550 nm of the patterned DA-AgNW networks with different sheet resistances.
}
    \label{fig:S14}
\end{figure}

\newpage
\begin{figure}[H] % H 表示强制当前位置插入
    \centering
    \includegraphics[width=\textwidth]{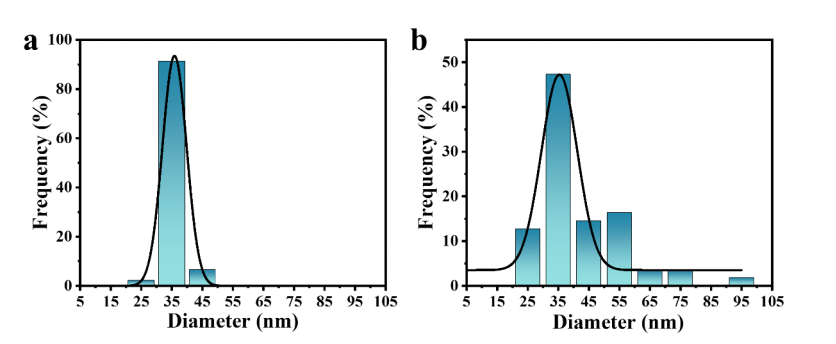} % 替换为你的图片文件
    \caption{\textbf{Fig. S15.} The distribution of AgNWs with different diameters (a) before and (b) after GTE-induced nanowire fracture. 
}
    \label{fig:S15}
\end{figure}

\begin{figure}[H] % H 表示强制当前位置插入
    \centering
    \includegraphics[width=\textwidth]{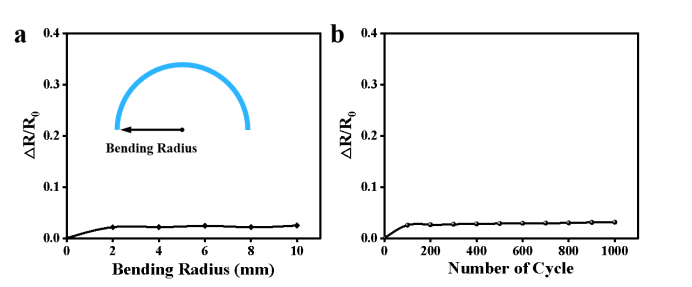} % 替换为你的图片文件
    \caption{ \textbf{Fig. S16.} Bending stability of DA-AgNWs. (a) \( \Delta R / R_0 \) as a function of bending radius (\(\Delta R\) and \( R_0 \) represent the resistance change and the initial resistance, respectively). (b) \( \Delta R / R_0 \) as a function of the number of cycles of repeated bending to a radius of 4 mm.
}
    \label{fig:S16}
\end{figure}

\newpage
\begin{figure}[H] % H 表示强制当前位置插入
    \centering
    \includegraphics[width=\textwidth]{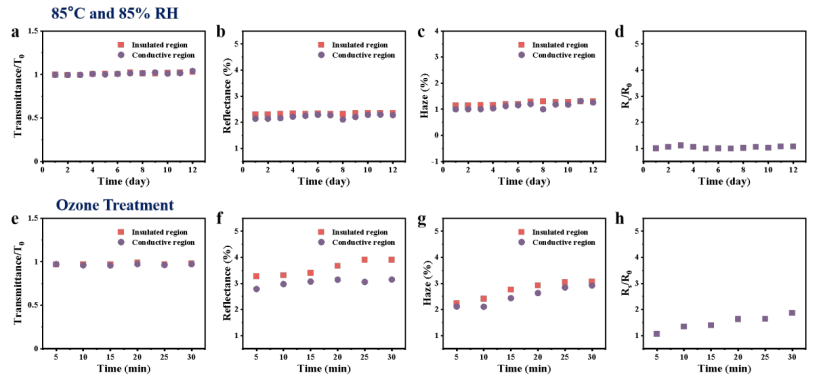} % 替换为你的图片文件
    \caption{\textbf{Fig. S17.} Oxidation resistance of DA-AgNWs. (a) - (d) Changes in haze, transmittance, reflectance, and \( R_S \) of the DA-AgNW sample at 85\(^\circ\)C and 85\% RH for 12 days. (e) - (h) Changes in haze, transmittance, reflectance, and \( R_S \) during 30-min ozone treatment.
}
    \label{fig:S17}
\end{figure}

\begin{figure}[H] % H 表示强制当前位置插入
    \centering
    \includegraphics[width=\textwidth]{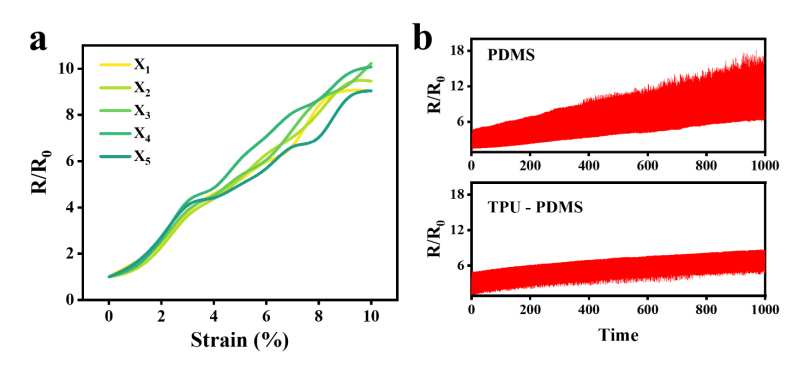} % 替换为你的图片文件
    \caption{\textbf{Fig. S18.} Electromechanical properties of AgNW electrodes for the touch sensor. (a) Changes in \( R/R_0 \) during stretching. (b) \( R/R_0 \) variations of the AgNW/PDMS electrode and TPU-encapsulated AgNW/PDMS electrode under a 1000-cycle stretching-releasing test with a maximal strain of 5\%.
}
    \label{fig:S18}
\end{figure}

\begin{figure}[H] % H 表示强制当前位置插入
    \centering
    \includegraphics[width=\textwidth]{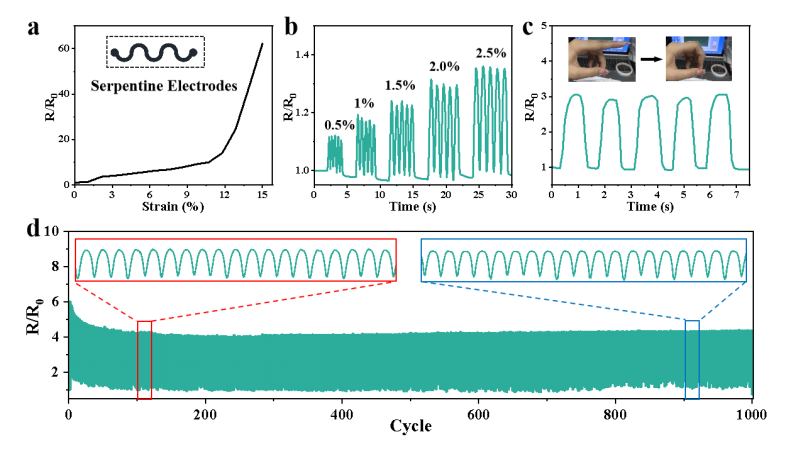} % 替换为你的图片文件
    \caption{\textbf{Fig. S19.} Performance of strain sensors based on the DA-AgNW patterns. (a) \( R/R_0 \) change under 0–15\% stretching. (b) \( R/R_0 \) variation of the strain sensor under small strains. (c) Detection response of the strain sensor to index finger bending. (d) \( R/R_0 \) variation of the strain sensor under 5\% strain for 1000 cycles.}

    \label{fig:S19}
\end{figure}

\newpage

\begin{table}[ht]
    \centering
    \caption{Table S1. Comparison of the GTE method with other traditional patterning methods}
    \label{tab:S1}
    \renewcommand{\arraystretch}{1.2} % 调整行距

    % 自动缩小表格
    \resizebox{\textwidth}{!}{  
    \begin{tabular}{|l|cccc|ccc|c|}
        \hline
        \multirow{2}{*}{\textbf{Methods}} & \multicolumn{4}{c|}{\textbf{Steps}} & \multicolumn{3}{c|}{\textbf{Performance}} & \textbf{Optical} \\ 
        \cline{2-8}
        & Coating & Exposure & Heating & Extra & $\Delta H\%$ & $\Delta T\%$ & Linewidth ($\mu$m) & \textbf{invisibility} \\ 
        \hline
        Ink printing$^{2}$ & \checkmark & - & - & - & - & 20 & 70 & No \\ 
        \hline
        Ink printing$^{3}$ & \checkmark & - & 110$^\circ$C for 5 min & - & - & - & 100 & No \\ 
        \hline
        Ink printing$^{4}$ & \checkmark & - & 350$^\circ$C for 30 min & - & - & - & 21 & No \\ 
        \hline
        Photolithography$^{5}$ & \checkmark & \checkmark & 120$^\circ$C for 5 min, 100$^\circ$C for 90 s & Development, Etching, Photoresist stripping & - & - & 20 & No \\ 
        \hline
        Wax-shaped wetting/dewetting$^{6}$ & \checkmark & - & 120$^\circ$C for 30 min, 150$^\circ$C for 15 min & Lithography, Wax microfilm removal & - & 21 & 60 & No \\ 
        \hline
        Ultrasonication$^{7}$ & \checkmark & \checkmark & - & Ultrasonication, Releasing & - & 10 & 30 & No \\ 
        \hline
        PRI$^{8}$ & \checkmark & \checkmark & 193$^\circ$C for 10 min & Development, Photoresist stripping & 0.8 & 1.6 & 30 & Yes \\ 
        \hline
        GTE & \checkmark & \checkmark & 75$^\circ$C for 3 min & - & 0.3 & 1.4 & 10 & Yes \\ 
        \hline
    \end{tabular}
} % 结束 resizebox
\end{table}

\newpage
\textbf{Reference}

1.	Yang, H. U. et al. Optical dielectric function of silver. Phys. Rev. B 91, 235137 (2015).

2.	Mazzotta, A. et al. Invisible thermoplasmonic indium tin oxide nanoparticle ink for anti-counterfeiting applications. ACS Appl. Mater. Interfaces 14, 35276-35286 (2022).

3.	Li, W. et al. Microsecond-scale transient thermal sensing enabled by flexible Mo1 - xWxS2 alloys. Research 7, 0452 (2024).

4.	Li, W. et al. Large‐scale ultra‐robust MoS2 patterns directly synthesized on polymer substrate for flexible sensing electronics. Adv. Mater. 35, 2207447 (2023).

5.	Ahn, Y., Lee, H., Lee, D. \& Lee, Y. Highly Conductive and Flexible Silver Nanowire-Based Microelectrodes on Biocompatible Hydrogel. ACS Appl. Mater. Interfaces 6, 18401-18407, doi:10.1021/am504462f (2014).

6.	Ma, P. et al. Wax-shaped wettability assisted patterning of silver nanowires on various substrates as transparent, flexible, or stretchable electrodes. Appl. Surf. Sci. 639, 158232 (2023).

7.	Liu, G.-S. et al. Ultrasonically Patterning Silver Nanowire–Acrylate Composite for Highly Sensitive and Transparent Strain Sensors Based on Parallel Cracks. ACS Appl. Mater. Interfaces 12, 47729-47738, doi:10.1021/acsami.0c11815 (2020).

8.	Liu, G.-S. et al. Self-assembled monolayer modulated Plateau-Rayleigh instability and enhanced chemical stability of silver nanowire for invisibly patterned, stable transparent electrodes. Nano Res. 15, 4552-4562 (2022).